\documentclass[twocolumn,pra]{revtex4}

\usepackage{ORI_Group_style}

\begin{document}

\title{Near-field Levitated Quantum Optomechanics with Nanodiamonds}
  
\author{M. L. Juan$^{1,2}$}
\email{mathieu.juan@mq.edu.au}
\author{G. Molina-Terriza$^{1,2}$}
\author{T. Volz$^{1,2}$}
\author{O. Romero-Isart$^{3,4}$}

\affiliation{$^1$ Department of Physics \& Astronomy, Macquarie University, NSW 2109, Australia}
\affiliation{$^2$ ARC Centre for Engineered Quantum Systems, Macquarie University, NSW 2109, Australia}
\affiliation{$^{3}$Institute for Quantum Optics and Quantum Information of the
Austrian Academy of Sciences, A-6020 Innsbruck, Austria.}
\affiliation{$^{4}$Institute for Theoretical Physics, University of Innsbruck, A-6020 Innsbruck, Austria.}

\begin{abstract}
We theoretically show that the dipole force of an ensemble of quantum emitters embedded in a dielectric nanosphere can be exploited to achieve near-field optical levitation. The key ingredient is that the polarizability from the ensemble of embedded quantum emitters can be larger than the bulk polarizability of the sphere, thereby enabling the use of repulsive optical potentials and consequently the levitation using optical near-fields. In levitated cavity quantum optomechanics, this could be used to boost the single-photon coupling by combining larger polarizability to mass ratio, larger field gradients, and smaller cavity volumes while remaining in the resolved sideband regime and at room temperature. A case study is done with a nanodiamond containing a high-density of silicon-vacancy color centers that is optically levitated in the evanescent field of a tappered nano-fiber and coupled to a high-finesse microsphere cavity.
  
\end{abstract}

\maketitle

\section{Introduction}
\label{intro}

Both a two level atom and a lump of dielectric material are polarizable objects whose motion can be manipulated with optical light~\cite{AshkinBook}, a feature that has been exploited in the fields of ultracold atoms~\cite{ReviewUCA} and optical tweezers~\cite{ReviewOT}. A recent experiment~\cite{MacquarieTrapping} has shown that the force from quantum emitters embedded in a dielectric nanosphere can be observed--even in liquid. In this article we go a step further and argue that it is experimentally feasible to have a scenario where the optical dipole force due to the embedded quantum emitters is {\em stronger} that the bulk dipole force. This would allow to use repulsive optical potentials for dielectric nanospheres, by using the internal structure of the quantum emitters, to trap them in evanescent fields analogously to what can be done with ultracold atoms~\cite{Vetsch2010}. We propose to use this for dispersive cavity quantum optomechanics~\cite{ReviewOpto} with optically levitated nanospheres~\cite{RomeroIsart2010,Chang2010,Barker2010,RomeroIsart2011,Kiesel2013,Millen2014,Gieseler2012}.
In addition we show that it might be possible to reach the so-called strong single-photon cooperativity regime~\cite{ReviewOpto,Mancini1997,Bose1997,Ludwig2008,Rabl2011,Nunnenkamp2011,Qian2012}, in the resolved sideband regime~\cite{ReviewOpto}, by coupling the levitated nanospheres to a high-finesse optical micro cavity, at room temperature.  

In dispersive cavity quantum optomechanics~\cite{ReviewOpto}, the single photon coupling $g_0$ is proportional to $\alpha b/(V_c M^{1/2})$, where $\alpha$ is the real part of the polarizibility, $b$ is the gradient of the cavity field mode, $V_c$ is the cavity volume and $M$ the mass of the mechanical oscillator. Our proposal aims at simultaneously combining salient features that have been demonstrated in independent experiments: i) placing a dielectric nanobject in the near-field of a microcavity to have larger $b$ and smaller $V_c$~\cite{Anetsberger2009}, ii) use a set of $N$ quantum emitters to have larger $\alpha/M^{1/2} \propto \sqrt{N}$~\cite{Murch2008,Brennecke2008}, and iii) optically levitate a nanosphere in high vacuum to have a high mechanical quality factor at room temperature~\cite{Gieseler2012} with a sufficiently large trap frequency to enable cavity cooling in the resolved sideband regime~\cite{Kiesel2013}. This conjunction of  features could be used for observing non-Gaussian physics of mechanical nanooscillators~\cite{Mancini1997,Bose1997,Ludwig2008,Rabl2011,Nunnenkamp2011,Qian2012},  measuring short-distance forces~\cite{Geraci2010}, and migrating to  quantum optomechanics cutting-edge experiments and proposals done with ultracold atoms in near-fields~\cite{Alton2011,Reitz2013,Thompson2013,Petersen2014,Chang2013,Parkins2014, Goban2014, Chang2014, GonzalezTudela2014, Ramos2014}. We remark that the increase of the polarizability of diamond by doping it with a high-density of color centers might also be used to boost the optomechanical coupling in setups using clamped cantilevers, membranes, or photonic crystals made of diamond~\cite{Ovartchaiyapong2014, Teissier2014, Kipfstuhl2014}. See~\cite{Dantan2014} for similar ideas.

The paper is organized as follows. In Sec.~\ref{pola} we discuss and compare the polarizability of quantum emitters to the bulk polarizability of the nanosphere. We show that for a particular type of color centers in diamond, the polarizability of the quantum emitters can overcome the bulk polarizability. Assuming this regime, we describe in Sec.~\ref{trap} the optical trap obtained using a bi-chromatic field supported by a nano-fiber. In Sec.~\ref{cavity_optomecha} we show that by placing a micro-cavity in close proximity to the nanosphere it is possible to achieve both the resolved sideband and the strong single-photon quantum optomechanical cooperativity regime. We conclude in Sec.~\ref{conclusion}.

\section{Quantum emitter polarizability}
\label{pola}

\subsection{General expression}

Let us compare the polarizability of a quantum emitter (called quantum polarizability hereafter) with the bulk polarizability of a dielectric nanosphere. In general terms, the time-averaged dipole force describing the interaction of a monochromatic field $\EE(\rr,t)$ of frequency $\w$ with a particle of polarizability $\alpha$ is given by~\cite{AshkinBook}:
\be
\FF(\rr) =  \alpha E_0(\rr) \frac{\grad E_0(\rr)}{2},
\ee
where $E^2_0(\rr)=2 \avg{\abs{\EE(\rr,t)}^2}$ (here $\avg{\cdot}$ denotes time average). In the case of a bulk dielectric nanosphere of refractive index $n$ and radius $R$, the polarizability $\alpha_s$ is given in the point dipole approximation by
\be
\alpha_s = 3 \epsilon_0 V \frac{n^2-1}{n^2+2},
\ee
where $\epsilon_0$ is the vacuum permittivity and $V=4 \pi R^3/3$. The validity of the point dipole approximation, usually assumed by the condition $R\ll 2\pi c / \omega$ for a focused Gaussian beam, can be verified in the context of near-fields  using a multimodal decomposition for the near-field~\cite{Barchiesi1993,Zvyagin1998} and the dielectric sphere. Using typical experimental parameters one can show that the interaction can be very well approximated by the point dipole term.

For a two-level state quantum emitter with transition frequency $\w_0$, dipole moment $d$, Rabi frequency $\Omega = \sqrt{2} \abs{\bra{e}  \dd \cdot \vec \epsilon \ket{g}  } E_0(\rr)/\hbar$, spontaneous emission in free space $\Gamma_0 = d^2 \w_0^3/(3 \epsilon_0 \pi \hbar c^3)$ {(accounting for an orientational average for the dipole moment $\abs{\bra{e}  \dd \cdot \vec \epsilon \ket{g}  }^2=d^2/3$)}, spontaneous emission inside the dielectric nanosphere $\Gamma \approx n \Gamma_0$~\cite{Inam2011}, and transverse decay rate $\gamma = \Gamma/2 + \gamma_c$, where $\gamma_c$ accounts for the additional coherence decay (inhomogeneous broadening), the quantum polarizability $\alpha_q$ is given by~\cite{Steck}
\be
\alpha_q=- \frac{2 \Delta  d^2 \Gamma}{3 \hbar \Omega^2 \gamma} \frac{s}{s+1}.
\ee
Here $\Delta \equiv \w - \w_0$ is the detuning and $s\equiv \Omega^2/[\gamma \Gamma (1+\Delta^2/\gamma^2)]$ the saturation parameter. This description uses the rotating wave approximation, which is valid provided $\abs{\Delta} \ll \w_0 $, and the Born-Oppenheimer approximation, which is valid provided the motional dynamics are much slower than the electronic dynamics of the quantum emitter. Note that $\alpha_q$ can be maximized to $\alpha_q =  - d^2/(3 \hbar \Delta)$ for the optimal detuning $|\Delta|= \gamma [1+ \Omega^2/(\gamma \Gamma)]^{1/2}$ that leads to a saturation parameter $s  = \Omega^2 (\Gamma \gamma)^{-1} (1+\Delta^2/\gamma^2)^{-1} \leq 1$. The optimal ratio between the two polarizabilities $\eta \equiv |\alpha_q/\alpha_s |$ can thus be written as
\be
\eta  =  \frac{ \lambda^3}{R^3} \frac{  2 }{ (4 \pi)^3}     \frac{n^2 +2}{n^2-1} \frac{\Gamma}{ n \gamma }  \frac{1}{ \sqrt{1+ \Omega^2/(\gamma \Gamma)}} .
\ee
Assuming that in the nanosphere one has a number of embedded identical quantum emitters given by $N = \rho_q V$, where $\rho_q$ is the volume density of emitters, and that the variation of fields within the nanosphere is negligible such that each quantum emitter identically interacts with the field (point-particle approximation), the total quantum polarizability will be larger than the bulk polarizability when $N \eta  > 1$. {Alternatively, an effective complex refractive index $\bar{n}$ can be defined from the complex polarizability $\bar{\alpha_q}$ of the ensemble of emitters by using the Lorentz-Lorenz relation $\rho_q \bar{\alpha_q}/(3\epsilon_0)=(\bar{n}^2-1)/(\bar{n}^2+2)$ for which one would obtain a value close to $2i$ near the resonance of the quantum emitters.} {It is important to note that we have not taken into account cooperative effects. These effects could have a significant impact on the dipole force and consequently polarisability ratio $\eta$}.

\begin{figure}[t]
  \centering
  \includegraphics[width= \columnwidth]{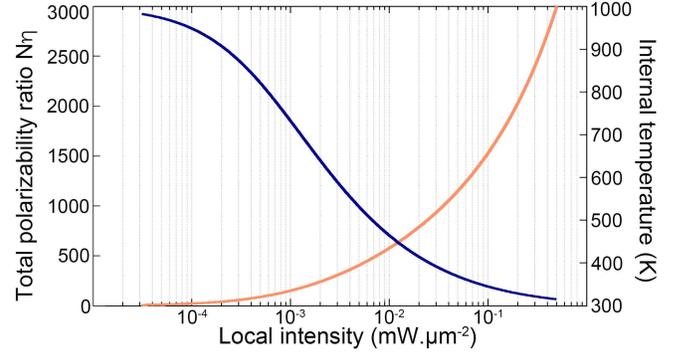} 
  \caption{Ratio $N \eta$ (solid dark line) of the total quantum polarizability over the dielectric polarizability for a nano-diamond with a high density of SiV. Larger intensities both increase the saturation of the quantum emitters and the internal temperature (solid light line) leading to a reduction of the quantum polarizability. The properties of the nanodiamond and the SiV are listed in Appendix~\ref{param_diamond} and \ref{param_siv} respectively. We remark that since the SiV centers are embedded in the ND, we accounted for the Lorentz local field correction~\cite{Maki1991}, $(n^2+2)/3$, for the field inside the ND. } \label{Fig1}
\end{figure}

\subsection{Color centers in diamond: silicon-vacancy centers}

Let us consider a levitated nanodiamond (ND)~\cite{Neukirch2013,Neukirch2015,Hoang2016} for its remarkable optical properties and the large variety of color centers they have~\cite{Jelezko2006,Aharonovich2011}, which act as quantum emitters. In particular, the silicon-vacancy (SiV) centers, consisting of a silicon atom and two adjacent vacancies, have the following favorable properties for achieving $N\eta \gg 1$: i) strong dipole moment~\cite{Vlasov2014}, ii) high-densities~\cite{Vlasov2014}, and iii) very good indistinguishability of different SiV~\cite{Rogers2014}. 

{The polarizability of the quantum emitters varies linearly with the intensity of the incident field for small intensities ($\Omega^2 \ll \gamma\Gamma$) leading to a constant ratio $\eta$. For larger intensity, this ratio decreases due to saturation effects from the quantum emitters. The incident intensity also has an impact through the dependence of the emitters with the internal temperature of the ND~\cite{Jahnke2014} (see Appendix~\ref{param_siv})}. In the context of levitation, this effect is important as the ND can reach high temperatures even with low intensities, see \figref{Fig1}. In order to account for this effect, the internal energy of the ND is obtained as a function of the absorption rate of the incident laser photons, $\gamma_{l}$, the absorption rate of the black body radiation of the environment, $\gamma_{a}$, and the black body emission rate, $\gamma_{e}$, is given by~\cite{Bateman2014}
\be\label{eq:rate_balance}
\begin{split}
m c_m \dot T_{i} = \gamma_{l} + \int \text{d}\omega' \left[ \gamma_{a} \left( \omega' \right)- \gamma_{e}\left( \omega', T_{i} \right) \right] \hbar \omega',
\end{split}
\ee
where $T_{i}$ is the internal temperature of the ND, $m$ its mass, and $c_m$ its specific heat.
For an incident laser of intensity $I$ and frequency $\omega$, and  an environmental temperature $T_{e}$, these rates are given by
\bea
\gamma_{l} &=& \frac{4 \pi I \omega R^3}{c} \text{Im}\left[ \frac{\epsilon(\omega) - 1 }{\epsilon(\omega) + 2}\right],\\
\gamma_{a} &=& \frac{4 \left( \omega' R / c\right)^3/ \pi}{\text{exp} \left( \hbar \omega' / K_b T_{env} \right) - 1} \text{Im}\left[ \frac{\epsilon(\omega') - 1 }{\epsilon(\omega') + 2}\right],\\
\gamma_{e} &= & \frac{4}{\pi} \left( \frac{\omega' R}{c} \right)^3 \text{exp} \left( -\frac{\hbar \omega'}{K_b T_{i}} \right) \times \text{Im}\left[ \frac{\epsilon(\omega') - 1 }{\epsilon(\omega') + 2}\right].\label{Eq:emBB}
\eea
The steady state temperature ($\dot T_{i}=0$) is determined as a function of the incident laser intensity (see Fig.~\ref{Fig1}) using the rate-balance equation for the internal energy (\eqcite{eq:rate_balance}). Accounting for the temperature dependence of the linewidth, lifetime and transition of the SiV centers, see App.~\ref{param_siv}, the total quantum polarizability for a ND containing a large density of centers clearly overcomes the bulk polarizability for low intensities (see Fig.~\ref{Fig1}). {The internal temperature will be largely modified by the non-radiative properties of the embedded quantum emitters which we have not accounted for here. In particular, the main contribution will be from the limited quantum efficiency of SiV centres and, to a lesser extend, from its emission in the phonon sidebands. In addition, it is important to note that the experimental values used to estimate the emissivity of the diamond (see Appendix~\ref{param_diamond}) are limited to the 2.5-6.5$\mu$m range leading to an underestimated black body emission rate. The internal temperature of the nano-diamond could be further reduced by coating its surface with a small layer of glass~\cite{Neukirch2015} providing much higher emissivity in the infra-red. By achieving much lower internal temperatures, the optical properties of the SiV centers will significantly improve.}

\section{Near-field optical trap}
\label{trap}

In the following we assume that the quantum polarizability dominates $N \eta \gg 1$ and discuss in a general way how this can be used to levitate a dielectric nanosphere in evanescent fields. In analogy to bichromatic atomic optical traps~\cite{Ovchinnikov1991, Kien2004, Sague2008}, we consider an evanescent field in vacuum of the form:
\be \label{eq:bichrom}
\EE(\rr, t) = E_1(\rr) \vec \epsilon_1 e^{-\im \w_1 t} + E_2(\rr) \vec \epsilon_2 e^{-\im \w_2 t+\phi} + \cc
\ee
The exponentially decaying electric field amplitudes are given by $E_{1(2)}(\rr) = E_f(x,y) e^{-\Lambda_{1(2)} z}$, where $\Lambda_{1(2)}$ is the field decay rate and $E_f(x,y)$ the field amplitude at the dielectric surface. The polarization of each mode is given by $\vec \epsilon_{1(2)}$ and $\phi$ is the phase difference between the two modes at $t=0$. {This field can be obtained from a nano-fiber and is different from the cavity field which will provide the back-action.} We consider the symmetric driving $\w_1 = \w_0 - \Delta$ and $\w_2 = \w_0 + \Delta$ with $\Delta>0$ and define the beating frequency $\delta = \w_2-\w_1 = 2 \Delta$. We are interested in the interaction of such field with a two-level quantum emitter when far-detuning or weak driving for the two modes is not assumed. This is an unusual scenario for most of the atomic trapping experiments with evanescent fields~\cite{Vetsch2010}; notwithstanding, this was studied before both experimentally and theoretically, see for instance~\cite{Grimm1990,Papademetriou1992,Ficek1993,Grimm1995,Ficek1996} and reference therein. One can analytically calculate the time-averaged dipole force considering that the beating is much faster than the dynamics of the mechanical motion. This is done solving a recursive equation obtained with a Floquet analysis. As shown below, one encounters that the total dipole force is not simply the sum of the forces that each mode would exert in the absence of the other mode but a more involved expression containing mixing terms that reflect the intricate interplay between the red and blue driving fields.

In particular, for a bichromatic field (\eqcite{eq:bichrom}), the Hamiltonian of the system in the rotating frame defined by the unitary $\hat U(t) = \exp \spare{ \im \w_1 \ketbra{e}{e} t}$ reads
\be
\Hop = \hbar \Delta \ketbra{e}{e} - \hbar \dd_{eg} \cdot \EE(\rr, t) e^{\im \w_1 t} \spl - \hc.
\ee
Here $\dd_{eg} = \bra{e} \dd \ket{g}$ and $\rr$ is the position of the two-level system. Applying the rotating wave approximation (valid provided $\Delta \ll \w_0$) this Hamiltonian can be written as
\be
\Hop = - \hbar \Delta \ketbra{e}{e} + \frac{\hbar \Omega_1(\rr)}{2} \spare{ 1  + \Xi(\rr) e^{-\im  \delta t} } \spl + \hc
\ee
where we have defined
\be
 \Omega_{1(2)}(\rr) 
 \equiv - \frac{2}{\hbar} \dd_{eg} \cdot \vec \epsilon_{1(2)} E_{1(2)}(\rr), 
 \ee
 which is also assumed to be real, and $\Xi(\rr) = \Omega_2(\rr)/\Omega_1(\rr)$. Hereafter we omit the $\rr$ dependence to ease the notation. 
Assuming that the internal electronic dynamics of the two-level system are much faster than the motional dynamics (Born-Oppenheimer approximation), the force is given by
\be \label{eq:tForce}
\FF = -  
\grad \spare{\frac{\hbar \Omega_1}{2}  \pare{1+ \Xi e^{-\im  \delta t} }} \avg{\spl} + \hc,
\ee
where the expected values are calculated for the electronic steady state.  Following~\cite{Papademetriou1992} the steady state solution can be expanded as a Fourier series in terms of $\delta$ (Floquet's analysis) such that
\bea
u(t) &\equiv&  \avg{\smi} = \sum_{n=-\infty}^\infty u_n e^{\im n \delta t}, \\
v(t) &\equiv&  \avg{\spl} = \sum_{n=-\infty}^\infty v_n e^{\im n\delta t}, \\
w(t) &\equiv&  \avg{\sz} = \sum_{n=-\infty}^\infty w_n e^{\im n\delta t}. 
\eea
The Fourier coefficients are obtained by solving the optical Bloch equations in the steady state, namely
 \bea \label{eq:uvw} \nonumber
 0 &=& - \spare{ \gamma + \im \delta (n+1/2)} u_n + \im \frac{\Omega_1}{2}  \pare{ w_{n}  + \Xi w_{n+1}},    \\  \nonumber
0&=& - \spare{\gamma+\im  \delta (n-1/2) } v_n -  \im \frac{\Omega_1}{2}  \pare{ w_{n}  + \Xi w_{n-1} } ,   \\  \nonumber
0 &=& -  (\Gamma + \im n \delta)w_n - \im \Omega_1  \pare{ v_{n}  + \Xi v_{n+1} } \\
 && + \im \Omega_1 \pare{u_{n}  + \Xi u_{n-1} }     - \Gamma \delta_{n0}.
\eea
Then, by further assuming that $\delta$ is much larger that the motional frequency (secular approximation), one can time-average  \eqcite{eq:tForce} for the steady state to obtain
 \be
 \FF = - \frac{\hbar \grad { \Omega_1}}{2} \pare{  v_0     + u_{0}     }- \frac{\hbar \grad \pare{\Omega_1\Xi}}{2} \pare{  v_{1}   +  u_{-1}   }.
  \ee
 Note that the time-averaged population of the excited state $p_e$ will be given by
 \be
p_{e} = \frac{ w_0 +1}{2} =  \frac{\im \Omega_1}{2\Gamma} \spare{ u_0 - v_0 + \Xi \pare{u_{-1} - v_{1}}}
 \ee
In addition, \eqcite{eq:uvw} leads to the following recursive equation for $w_n$
\be
a_n w_n+b_n w_{n+1} + c_n w_{n-1} = d_n,
\ee
where
\bea
a_n &=& - \Gamma - \im n \delta - \frac{4 \Omega_1^2 (\gamma + \im n \delta)(1+\Xi^2)}{4 \gamma^2+\im 8 n \gamma \delta + \delta^2(1-4 n^2)}, \\
b_n &=& - \frac{2 \Xi \Omega_1^2}{2 \gamma + \im \delta (2 n+1)}, \\
c_n &=& - \frac{2 \Xi \Omega_1^2}{2 \gamma + \im \delta (2 n-1)}, \\
d_n &=& \Gamma \delta_{n0}.
\eea
By obtaining the set of $w_n$ on can readily calculate $u_n$ and $v_n$. This recursive equation can be exactly solved by fixing a cutoff $\mathcal{N}\ge 0$ such that $w_{\mathcal{N}+1} = w_{-\mathcal{N}-1}=0$. The value of $\mathcal N$ is chosen such that the value of the physical quantities that are calculated converge.

In the lowest order in the Floquet analysis, namely for $\mathcal{N}=0$, one obtains:
\be
\FF_q = \hbar \Delta \frac{\Gamma}{2 \gamma} \frac{s_1 {\bf g}_1  - s_2 {\bf g}_2}{1 + s_1 + s_2}.
\ee
We defined $s_{1(2)} \equiv \Omega_{1(2)}^2 (\Gamma \gamma)^{-1} \spare{1+ \pare{\Delta/ \gamma}}^{-2}  $, 
${\bf g}_{1(2)} = \grad \log \Omega_{1(2)}(\rr) $, and $\Omega_{1(2)}(\rr) = 2 \abs{   \dd_{eg} \cdot \vec \epsilon_{1(2)} E_{1(2)}(\rr)}/\hbar$.
The population of the excited state of the quantum emitter is given by $p_e=[1-(1+s_1+s_2)^{-1}]/2$. The total optical force exerted on the nanosphere can then be estimated by using $\FF_T \approx N \FF_q + \FF_{s} $, with $\FF_s(\rr) =  \alpha_s E_0(\rr) \grad E_0(\rr)/2$ the optical dipole force due to the bulk polarizability.


\begin{figure}[t]
  \centering
  \includegraphics[width=  \columnwidth]{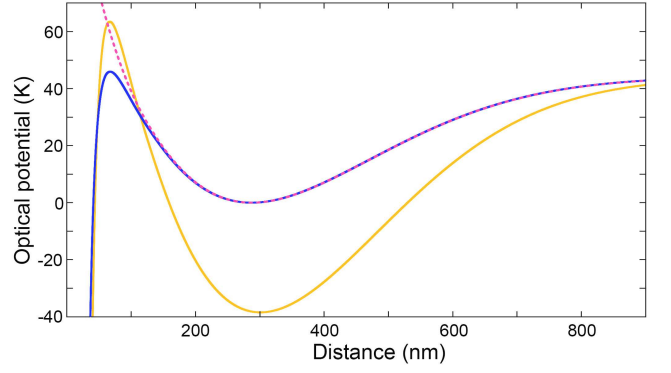} 
  \caption{Trapping potential for a 30~nm ND accounting for the dipolar forces acting on the quantum emitters and the ND along with gravity and the Casimir-Polder forces (see Appendix~\ref{param_fibre} for the experimental parameters). The trapping potential obtained using the dipole force for the bi-chromatic field using the second order in the Floquet analysis (solid line) shows an important deviation from considering the dipole force from the two evanescent fields independently (light solid line). For reference, the force without accounting for the Casimir-Polder forces is represented with the dashed line.} \label{Fig2}
\end{figure}

In order to achieve near-field levitation, the blue-detuned field ($\Delta > 0$) provides repulsive optical forces to prevent the nanosphere from adsorbing to the surface producing the near-fields. Consequently it is necessary to account for the Casimir-Polder force to fully capture the trapping potential. In order to simplify the calculation, we have approximated the surface as a semi-infinite half space of dielectric material with dielectric constant $\epsilon\left(\omega \right)$. The Casimir-Polder potential for an object of polarizability $\alpha(\omega)$ at a distance $z$ from the surface takes the form~\cite{Casimir1948, DispersionBook}:
\be
\begin{split}
& U^\text{cp}_{q(s)}(z) = \frac{\hbar}{8 c^2 \pi^2  \epsilon_0} \int_0^\infty \text{d} x x^2 \alpha_{q(s)}(\im x) \int_{x/c}^\infty \text{d}k e^{-2 k z} \times \\
& \times \spare{ \frac{k - g(x,k) }{ k+g(x,k)} + \pare{1-\frac{2 k^2 c^2}{x^2}} \frac{\epsilon(\im x) k - g(x,k) }{\epsilon(\im x) k + g(x,k)}},
\end{split}
\ee
where
\be
g(x,k) = \sqrt{ \frac{ x^2}{c^2} \spare{\epsilon(\im x) - 1} + k^2}.
\ee
For the polarizability of the quantum emitters with a transition frequency $\omega_0$ we use {$\alpha_q(\omega) = 2 d^2 \omega_0 [3 \hbar (\omega^2_0 - \omega^2)]^{-1}$}. For a dielectric sphere of refractive index $n$ and volume $V$ we use {$\alpha_s(\omega) =  3 \epsilon_0 V (n^2-1)/(n^2+2)$}. Considering silica for the semi-infinite half spaces, see Appendix and~\cite{Malitson1965}, one can numerically calculate the Casimir-Polder force. The total Casimir-Polder potential for a nano diamond containing $N$ quantum emitters is obtained as
\be
U_\text{cp}(z) = N U^\text{cp}_q(z) + U^\text{cp}_s(z).
\ee
Finally, one can then estimate the total force exerted on the nanosphere by using $\FF_T \approx N \FF_q + \FF_{cp} + \FF_{s} + \FF_{g}$. Here $\FF_{cp}=- \grad U_\text{cp}$ is the Casimir-Polder force and $\FF_{g}  $ is the gravitational force along the $z$-axis.

The most efficient near-field trap is obtained using the optimal detuning, $|\Delta|= \gamma [1+ \Omega^2/(\gamma \Gamma)]^{1/2}$, for the quantum emitters to maximize the polarizability ratio $\eta$. The trapping potential for the particular case of a 30~nm ND with embedded SiV centers is shown in  \figref{Fig2}. The possibility to levitate the nanosphere using near-fields, which provides much stronger optical forces, allows the use of low intensities and thereby the reduction of heating observed when using a focused Gaussian beam~\cite{Millen2014b}. The dephasing of the SiV, varying as the cube of the internal temperature~\cite{Jahnke2014}, remains sufficiently low to obtain a total quantum polarizability much larger than the bulk polarizability. On a practical level, the trap depth is particularly adapted to the use of mobile optical fiber traps as they provide a cooling of the center-of-mass motion down to 30~K with a relatively simple apparatus~\cite{Mestres2015}. The experimental parameters are described in Appendix~\ref{param_fibre}.

\section{Cavity optomechanics}\label{cavity_optomecha}

\subsection{Optomechanical coupling}

\begin{figure}[t]
  \centering
  \includegraphics[width=  \columnwidth]{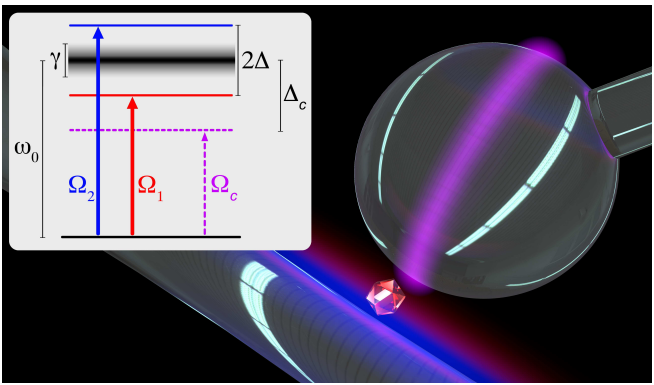} 
  \caption{Schematic illustration of the general scenario that is considered: a dielectric nanosphere with a set of two-level quantum emitters is trapped using a bichromatic evanescent field at a distance $z$ from a surface. In the inset of the figure the level structure of an individual quantum emitter of transition frequency $\w_0$ and excited state linewidth $\gamma$ is illustrated. The transition is driven by a bichromatic field with symmetric red and blue detuning $\mp \Delta > \gamma $, with Rabi frequencies $\Omega_1$ and $\Omega_2$, respectively. The cavity field mode with vacuum Rabi frequency $\Omega_c$ is detuned by $\Delta_c > \gamma$. } \label{Fig3}
\end{figure}

Let us now address the optomechanical coupling between the levitated nanosphere and an optical cavity mode (see \figref{Fig3}). This is achieved by placing the sphere in the evanescent field of an optical microcavity mode with frequency $\w_c$ and creation (annihilation) operator $\adop$ ($\aop$). The total Hamiltonian is then given by:
\be
\frac{\Hop}{\hbar} = \w_t \bdop \bop+\w_0 \ketbra{e}{e}  + \w_c \adop \aop+\frac{\Omega_c(\zop)}{2} \pare{\aop \spl + \adop \smi}.
\ee
Here we used the rotating wave approximation, which is valid provided $|\Delta_c|, \Omega_c \ll \w_0$, where $\omega_c - \omega_0 \equiv \Delta_c$.
The vacuum Rabi frequency is given by $\Omega_c(\zop) = \Omega_c \xi e^{- \Lambda_c \zop}$ with $\Omega_c = 2 d [\hbar \w_c/(3 V_c \epsilon_0)]^{1/2}/\hbar$ and $\Lambda_c$ the decay rate of the cavity's evanescent field. Here $V_c$ is the cavity mode volume, $\zop = z_{zp} (\bdop + \bop) $ is the displacement operator along the $z$-axis from the equilibrium position $z_t$, with $z_{zp}$ the zero point motion, and $\omega_t$ the mechanical frequency. The coefficient $\xi$ accounts for the particular mode shape and polarization type of the cavity such that $E_{cav} = \xi [\hbar \w_c/(V_c \epsilon_0)]^{1/2}$ is the vacuum field at the cavity/vacuum interface~\cite{Demchenko2013}. By further assuming $\abs{ \Delta_c} > \gamma \gg \Omega_c$ one can make a Schrieffer-Wolf transformation  to obtain
\be
\begin{split}
\frac{\Hop}{\hbar} = & \w_t \bdop \bop+\spare{\w_0 - \frac{\Omega_c^2(\zop)}{4\Delta_c} \adop \aop} \ketbra{e}{e}+ \w_c \adop \aop \\ & - \frac{\Omega_c^2(\zop)}{4 \Delta_c} \sz \adop \aop.
\end{split}
\ee
Since $\Omega_c^2/(4 \Delta_c) \ll \gamma$, the atomic level shift due to the cavity field will be irrelevant and thus can be safely dropped.  By Taylor expanding the vacuum Rabi frequency around the equilibrium position, one obtains
\be
\begin{split}
\frac{\Hop}{\hbar} = & \w_t \bdop \bop+ \w_0 \ketbra{e}{e}+ \w_c \adop \aop - g_0 \frac{\sz}{\w_0} \adop \aop (\bdop+\bop)
\end{split}
\ee
By considering that the single-photon coupling $g_0$ is much smaller {than} the frequency of the internal dynamics one replaces the $\sigma_z$ by the $\avg{\sigma_z}=w_0$ calculated in the steady state considering the bichromatic driving. Therefore we arrive at the final single-photon optomechanical coupling Hamiltonian
\be
\begin{split}
\frac{\Hop}{\hbar} = & \w_t \bdop \bop+ \w_0 \ketbra{e}{e}+ \w_c \adop \aop - g_0 \adop \aop (\bdop+\bop).
\end{split}
\ee
where the single-photon optomechanical coupling due to $N$ quantum emitters is given by:
\be
g_0^{q} =  - N (2 p_e-1)  \xi^2 e^{- 2  \Lambda_c z'} \frac{ \Omega_c^2}{2 \Delta_c}  \Lambda_c   z_{zp}. 
\label{g0}
\ee
Note that the coupling due to the quantum emitters would vanish should the quantum emitters be totally saturated ($p_e=1/2$), leaving only the contribution due to the bulk polarizability from the nanosphere~\cite{Anetsberger2009}.

\subsection{Resolved sideband regime}

 \begin{figure}[t]
  \centering
  \includegraphics[width=  \columnwidth]{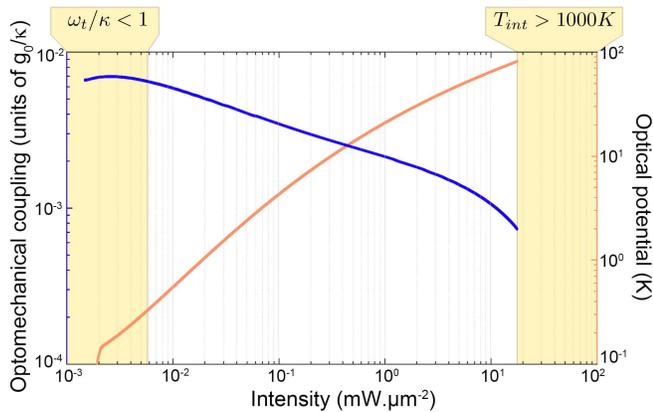} 
  \caption{Optomechanical coupling for a 30~nm ND while in the resolved sideband regime. The best optomechanical coupling is obtained at low-intensity as it provides a small excited state population (minimising the cavity scattering losses, see Eq.~\ref{kappaSC}) and low inhomogeneous broadening thanks to a moderate internal temperature (maximising the coupling, see Eq.~\ref{g0}). For the parameters see Appendix~\ref{param_sideband}.} \label{Fig4}
\end{figure}

In order to cool down the center-of-mass of the mechanical mode into the ground state using the optomechanical coupling to the cavity~\cite{Genes2008,Wilson-Rae2007,Marquardt2007} one requires the resolved sideband regime $\w_t \gtrsim \kappa$. In this context, the use of evanescent field provides much stronger gradient compared to focused Gaussian beams, allowing for larger mechanical frequencies. To verify that the resolved sideband regime can be reached, it is necessary to consider the impact that quantum emitters have on the optical quality factor through the scattering of cavity photons. For $N$ emitters, this is given by
\be
\kappa_{sc} = N \Gamma  \xi^2 e^{- 2  \Lambda_c z'} \frac{ \gamma \Omega_c^2 }{2(\gamma^2+\Delta_c^2)}.
\label{kappaSC}
\ee
Here, we assumed the saturation parameter from the cavity vacuum field is much smaller than 1, $s_{cav}= \xi^2\Omega_{c}^2 (\Gamma \gamma)^{-1} \spare{1+ \pare{\Delta/ \gamma}}^{-2} \ll 1$. In order to reduce this effect, we used a far detuned cavity, taking advantage of the excellent optical property of fused silica in the near infrared (see Appendix~\ref{param_cavity}). The cavity loss rate is then obtained as $\kappa = 2\pi c /(Q\lambda_{cav}) + \kappa_{sc}$.

By maintaining the ND away from the cavity surface, the impact of the embedded quantum emitter on the quality factor allows to maintain a strong optical trap in the resolved sideband regime with an optomechanical coupling comparable to the state of the art (see  Fig.~\ref{Fig4} and Appendix~\ref{param_sideband}). Upon reaching a low center-of-mass temperature, the coupling can be further increased by using lower intensities, \textit{i.e.} smaller dephasing (by maintaining a low internal temperature). As illustrated in Fig.~\ref{Fig4}, the coupling can be varied by one order of magnitude by changing the total intensity while remaining in the resolved sideband regime. Since the position of the nanodiamond is not changed, the optomechanical coupling decreases for large intensities due to the increase of both the internal temperature and the excited state population.

\begin{figure}[t]
  \centering
  \includegraphics[width=  \columnwidth]{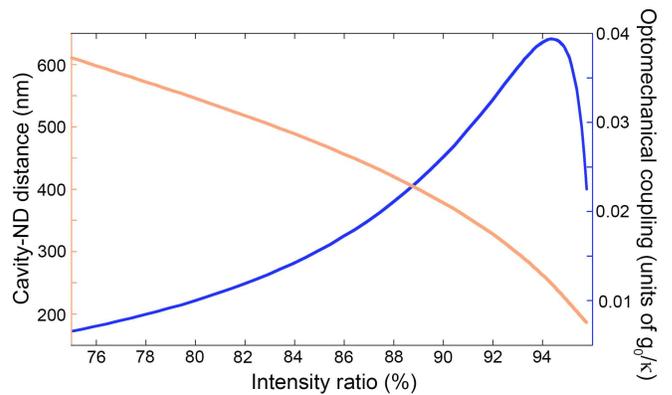} 
  \caption{Optomechanical coupling and ND position. By varying the intensity ratio $I_2/(I_1+I_2)$, the position of the 30~nm ND can be controlled in order to the reach the optimum optomechanical coupling. As the ND is brought close the cavity, the cavity scattering losses is greatly increased by the presence of the quantum emitters.} \label{Fig5}
\end{figure}

Another advantage of such bi-chromatic evanescent trap resides on the possibility to accurately control the position of the ND by simply varying the ratio of intensity between the two fields. Consequently, it is possible to further increase the optomechanical coupling by placing the ND closer to the cavity. \figref{Fig5} illustrates the evolution of the optomechanical coupling as a function of the intensity ratio. For each point, the total intensity is the lowest one still maintaining the resolved sideband regime. As the intensity of the repulsive field is increased, the ND is pushed further away from the nano-fiber, 
\textit{i.e.} closer to the cavity. This allows to increase the optomechanical coupling up until the cavity scattering losses reach values comparable to the unloaded cavity losses. For shorter cavity-ND distances, the ratio of the optomechanical coupling to the total scattering losses is reduced. In addition, higher intensities are required to achieve the resolved sideband regime, which increases the population of the excited state $p_e$ and hence reduces the coupling (\eqcite{g0}).

\subsection{Strong single-photon optomechanical cooperativity}

The challenging strong-single photon optomechanical coupling regime is achieved when $g_0 \gtrsim \kappa,\Gamma_m$~\cite{ReviewOpto}, where $\kappa$ is the decay rate of the optical cavity mode and $\Gamma_m$ is the decoherence of the center-of-mass mechanical motion. The leading contribution to $\Gamma_m$ in optically levitated nanospheres is the recoil heating due to scattering of photons~\cite{Chang2010,RomeroIsart2011,Pflanzer2012}. In our case there will be the additional and dominant contribution due to scattered photons from the quantum emitters. One can estimate that $\Gamma_m = \Gamma_m^q + \sum_{i=1}^2 \Gamma_m^{s,i}$, where the contribution from the quantum emitters is~\cite{Wineland1979,Cirac1992} $\Gamma_m^q= N (2/5) (\w_0 z_{zp}/c)^2  R_{sc}^q$ with the scattering rate from a single quantum emitter being $R^q_{sc}=\Gamma p_e$, and the contribution due to the bulk polarizability~\cite{Chang2010,RomeroIsart2011,Pflanzer2012} $\Gamma_m^s =\sum_{i=1}^2 (2/5) (\w_i z_{zp}/c)^2 R_{sc,i}^s$ with the scattering rate given by $R^s_{sc,i}=\abs{\alpha_s}^2 E_i^2(\rr_t) (\w_i/c)^3/(12 \pi \epsilon_0 \hbar)$ ($\rr_t$ is the trapping position of the sphere). {Although much weaker, the recoil due to the black body radiation from the nanosphere can be obtained from the black body emission rate Eq.~\ref{Eq:emBB} as $\Gamma_m^{BB} = \int\textnormal{d}\omega' [(2/5) (\omega' z_{zp}/c)^2 \gamma_e(\omega')]$.} One can now define the single-photon optomechanical cooperativity $\mathcal{C} = g_0^2/(\kappa \Gamma_m)$. We show below that by taking advantage of the quantum polarizability from quantum emitters embedded in the nanosphere, it is possible to obtain the strong single-photon optomechanical cooperativity regime $\mathcal{C}>1$ while maintaining the resolved sideband regime.

In order to reduce the impact the quantum emitters have on the cavity losses, large detunings can be used to trap the ND. This approach, inspired from the far-off-resonance trapping (FORT) for cold-atoms~\cite{Miller1993}, relies on the use of a far detuned near-field trap to reduce the excited state population $p_e$. In this case, one constrain remains: the force from the quantum emitters should remain larger than the bulk one in order to provide blue-detuned repulsive fields. Another consequence of this configuration is the important reduction of the overall forces. To compensate for this, we propose to use an additional blue-detuned field from the microsphere cavity. Such field plays the role of the red-detuned field from the nano-cavity in preventing the ND from falling on the cavity under the Casimir-Polder forces. By using a far-of-resonance blue-detuned field, one can achieve a much stronger field gradient, further improving the trapping efficiency. Despite the added complexity of using trapping fields from both the nano-fiber and the cavity, this configuration allows for maintaining the resolved sideband regime while significantly reducing the population of the excited state by using a large detuning. In particular, \figref{Fig6} shows the trapping potential (see Appendix~\ref{param_coop} for the experimental parameters). The 30~nm ND is then maintained in the near-field of the cavity allowing to reach both the resolved sideband regime and the single-photon strong cooperativity $\mathcal{C} \gtrsim 1$, see Appendix for the set of experimental parameters.

 \begin{figure}[t]
  \centering
  \includegraphics[width=  \columnwidth]{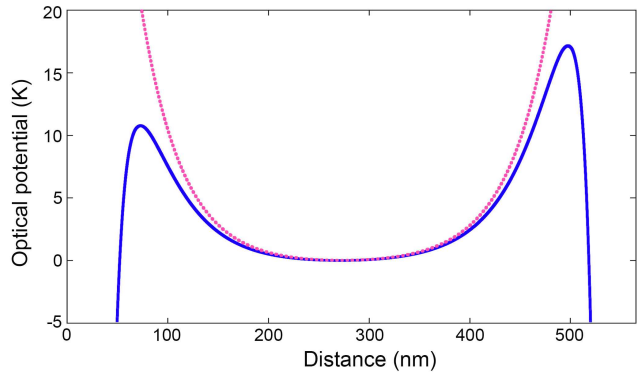} 
  \caption{ Trapping potential for a 30~nm ND using two blue-detuned fields. Two blue-detuned fields, one from the nano-fiber and from the cavity are used to trap the ND. The large detuning, $|\Delta|=1\times 10^{15}~\text{Hz}$, allows to maintain a very small excited state population (see Appendix~\ref{param_coop} for the experimental parameters). To provide confinement on the transversal dimensions, an additional weak red-detuned field from the nano-fiber can be used.} \label{Fig6}
\end{figure}

\section{Conclusions}
\label{conclusion}

In this article we have shown that the polarizability given by a set of quantum emitters embedded in a nanosphere can overcome its bulk polarizability. This immediately allows to use repulsive optical forces and levitate a nanosphere using evanescent fields. We have then discussed how this can be used for levitated quantum optomechanics using an evanescent coupling of the center of mass of the nanosphere with a microcavity mode. Due to a large polarizability to mass ratio, large cavity field mode gradient, and a smaller cavity mode volume, one could achieve the strong single photon cooperativity in the resolved sideband regime. A case study has be done considering a nanodiamond with a high density of embedded SiV color centers at room temperature. {Also, this work could be extended to a recently discovered color center in diamond, the germanium-vacancy center~\cite{Iwasaki2015}, as it provides significantly better optical properties at room temperatures.}

{We remark that as suggested in the recent experiments reported in~\cite{MacquarieTrapping}, collective effects from the high-density of quantum emitters embedded in the nanodiamond seem to be relevant. In particular, these effects could significantly modify the dipole force acting on the quantum emitters but also the total radiative emission of the emitters.} The latter is a key point since otherwise the imperfect quantum efficiency of single quantum emitters could heat up the nanodiamond to unfeasible temperatures. The theoretical understanding and potential applications of collective effects in this scenario, namely a levitated sphere smaller than the optical wavelength with such a high-density of quantum emitters that dipole-dipole interactions have to be taken into account~\cite{Friedberg1972,Gross1982} in addition to inhomogeneous broadening, is a very interesting and challenging further research direction that will be addressed elsewhere.

\section*{Acknowledgements}

ORI acknowledges support by the Austrian Federal Ministry of Science, Research, and Economy (BMWFW), and hospitality of EQuS and the QSciTech Distinguished Visitor Grant where this work was initiated.


\appendix

\section{Experimental parameters}
\label{param}

In the following we list all the experimental parameters needed to make the analysis of the experimental feasibility. We indicate the free parameters with a star symbol.

\subsection{Nanodiamond}\label{param_diamond}

\begin{itemize}

\item[$\star$]  Radius nanosphere: $R=15 ~\text{nm}$.

\item[$\star$]  Density of quantum emitters: $\rho_q = 1.4~\text{nm}^{-3}$. This density is an upper value extracted from the experimental measurements in \cite{Vlasov2014}, samples with smaller SiV densities would also fulfill $N\eta \gg 1$. 

\item Number of quantum emitters: $N=1.98 \times 10^4$.

\item Real part of the refractive index: $\text{Re}[n]=2.4$~\cite{Palik}. For the blackbody radiation, absorption and emission, the leading term in the integral for moderate temperature ($T_{int}, T_{env} \lesssim 2000$) arises from low energy radiation ($\omega \lesssim 2\times 10^{15} \text{Hz}$). This constant value for the real part of the refractive index of diamond was consequently used.

\item Imaginary part of the refractive index in the spectral region 500-1100~nm: $\text{Im}[n]=10^{-8}$. This value is estimated from the values of extinction available in the literature ($3.8\times10^{-7}$ at 436~nm~\cite{Palik}, $\lesssim2\times 10^{-8}$ at 532~nm and $\lesssim9\times10^{-9}$ at 1064~nm~\cite{Friel2010}).

The spectral dependence of the imaginary part of the refraction used is given by:
\be
\begin{split}
\text{Im}\spare{n(x)} = & 10^{-8}+  \\
& + \inv{4 \pi x}\spare{f_1(x) + f_2(x) + f_3(x)},
\end{split}
\ee
where
\bea
f_1(x)&=& 15 \exp\spare{- \pare{\frac{2100-x}{2 \times 10^5}}}, \\
f_2(x)&=& 1.5 \exp\spare{- \pare{\frac{3200-x}{2 \times 10^5}}}, \\
f_3(x)&=& 0.35 \exp\spare{- \pare{\frac{4400-x}{2 \times 10^5}}},
\eea
and $x \equiv \w/(100 \times 2 \pi c)$. Such function provides a satisfactory fit of the experimentally measured values in the wavelength range 3-6.5$\mu$m~\cite{Palik}. {In the absence of values for the extinction coefficient in the 6.5-20$\mu$m, we assumed a value of $10^{-8}$ which leads to a significant underestimation of the black body radiation of the nanodiamond.} The complex dielectric function of diamond $\epsilon\left(\omega\right)$ was obtained using this fit.


\end{itemize}

\subsection{Quantum emitter: silicon-vacancy}\label{param_siv}

\begin{itemize}

\item Transition frequency: $\w_0/2 \pi = 4.01 \times 10^{14}~\text{Hz}$.

Its bulk temperature's dependence is~\cite{Jahnke2014}:
\be
\frac{\lambda_0(T)}{1~\text{nm}} = 737 + 19.2 \times 10^{-8} \pare{\frac{T}{1~\text{K}}}^{2.78},
\ee
where $\lambda_0 =  2 \pi c/\w_0$.


Its bulk temperature's dependence is~\cite{Jahnke2014}:
\be
\begin{split}
\frac{\Gamma(T)}{2\pi\times 10^9~\text{Hz} }=& \inv{9.74}  \spare{1+3.3 e^{- 55~\text{meV}/(K_b T)}} 
\end{split}
\ee


Its bulk temperature's dependence is~\cite{Jahnke2014}:
\be
\frac{\gamma(T)}{2\pi \times 10^6~\text{Hz} }=16.39+1.9 \times 10^{-2} \pare{\frac{T}{1~\text{K}}}^3 
\ee


\end{itemize}

\subsection{Near-field trapping}\label{param_fibre}
\begin{itemize}

\item[$\star$] Nano-diamond size: 30~nm.

\item[$\star$] Fiber diameter: 715~nm.

\item[$\star$] Detuning: $\Delta$/2$\pi$ = $\pm$ 1$\times$10$^{13}$Hz.

\item[$\star$] Spatial decay of mode 1 (mode EH$_{21}$): $\Lambda^{-1} = $210~nm.

\item[$\star$] Intensity for mode 1: $I_1$ = 0.62mW/$\mu^2$.

\item[$\star$] Spatial decay of mode 2 (mode HE$_{11}$): $\Lambda^{-2} = $135~nm.

\item[$\star$] Intensity for mode 2: $I_2$ = 1.85mW/$\mu^2$.

\item Trap depth: $43~\text{K}$.

\item Trapping distance from the fiber's surface: \\ $z_t = 287~\text{nm}$.

\item Nanodiamond internal temperature: $587~\text{K}$.

\end{itemize}

\subsection{Microsphere cavity}\label{param_cavity}

\begin{itemize}

\item[$\star$]  Frequency of the cavity mode: $\w_c = \w_0 + \Delta_c$ where $\Delta_c = 1.4 \times 10^{15}~\text{Hz}$.

\item[$\star$]  Radius of the microsphere cavity:  $25~\mu\text{m}$.

\item Cavity intrinsic radiative losses: $Q_{rad}^{-1}$ = 2.2 $\times 10^{18}$.

\item Cavity scattering losses: $Q_{s.s.}^{-1}$ = 6.6 $\times 10^{18}$~\cite{Buck2003}.

\item Cavity material losses: $Q_{mat}^{-1}$ = 9 $\times 10^{10}$~\cite{Buck2003}.

\item[$\star$]  Cavity mode quality factor: $Q= 10^{10} < 1/(Q_{s.s.}^{-1} + Q_{bulk}^{-1} + Q_{rad}^{-1}) $.

\item  Decay rate of the cavity mode: \\ $\kappa/2 \pi = \w_c /(2\pi Q) = 18.3 \times 10^3 ~\text{Hz}$.

\item Cavity volume: $V_c=820~\mu\text{m}^3$.

\item  Decay of the evanescent field of the cavity mode: $1/\Lambda_c=283 ~\text{nm}$.

\end{itemize}

\subsection{Resolved side-band regime}\label{param_sideband}

\begin{itemize}

\item[$\star$] Distance between the nano-fiber and the cavity:  $D = 900~\text{nm}$.

\item[$\star$] Detuning: $\Delta$/2$\pi$ = $\pm$ 1$\times$10$^{13}$Hz.

\item[$\star$] Spatial decay of mode 1 (mode EH$_{21}$): $\Lambda^{-1} = $210~nm.

\item[$\star$] Intensity for mode 1: $I_1$ = 0.62mW/$\mu^2$.

\item[$\star$] Spatial decay of mode 2 (mode HE$_{11}$): $\Lambda^{-2} = $135~nm.

\item[$\star$] Intensity for mode 2: $I_2$ = 1.85mW/$\mu^2$.

\item Trap depth: $34~\text{K}$.

\item Trapping distance from the cavity: \\ $z_t = 612~\text{nm}$.

\item Nanodiamond internal temperature: $587~\text{K}$. 

\item Cavity loss rate due to the scattering from the emitters: $\kappa_{sc}/\kappa = 0.34$.

\item Mechanical decoherence rate due to photon scattering due to the emitters: $\Gamma_m^q/2 \pi=42.3 \times 10^{3}~\text{Hz}$.

\item Mechanical decoherence rate due to photon scattering due to the nanosphere: $\Gamma_m^s/2 \pi=0.63~\text{Hz}$.

\item Single-photon coupling vs. cavity loss rate: $g_0/\kappa=1.2 \times 10^{-3}$. 

\item Mechanical frequency: $\w_t/1\pi=1\times 10^{5}~\text{Hz}$.

\item Mechanical frequency vs. cavity loss rate: $\w_t/\kappa=4$.

\end{itemize}

\subsection{Strong single-photon optomechanical cooperativity (FORT)}\label{param_coop}

\begin{itemize}

\item[$\star$] Distance between the nano-fiber and the cavity:  $D = 565~\text{nm}$.

\item[$\star$] Far-of-resonance detuning: $|\Delta| = 1 \times 10^{15} ~\text{Hz}$.

\item[$\star$] Spatial decay of mode 2 (mode HE$_{11}$): $\Lambda^{-2} = $81~nm.

\item[$\star$] Intensity for mode 2: $I_2$ = 1.43mW/$\mu^2$.

\item Spatial decay of mode 3 (cavity mode): $\Lambda^{-3} = $85~nm.

\item[$\star$] Intensity for mode 3: $I_3$ = 1.72mW/$\mu^2$.

\item Trap depth: $10.8~\text{K}$.

\item Trap frequency: $\w_t/(2 \pi)=32.7 \times 10^3~\text{Hz}$.

\item Trapping distance from the fiber's surface: \\ $z_t = 270~\text{nm}$.

\item  Trapping distance from the microsphere cavity's surface: $z_t'=(565-270)~\text{nm}$.

\item Steady excited state population: $p_e=3.2\times 10^{-5}$.

\item Nanodiamond internal temperature: $T=385~\text{K}$.

\item Mechanical frequency vs. cavity loss rate: $\w_t/\kappa=1$.

\item Single-photon coupling vs. cavity loss rate: $g_0/\kappa=2.8 \times 10^{-2}$. 

\item Single-photon cooperativity: $\mathcal{C}=1.2$.

\item Cavity loss rate due to the scattering from the emitters: $\kappa_{sc}/\kappa = 0.79$.

\item Mechanical decoherence rate due to photon scattering from the emitters: $\Gamma_m^q/2 \pi=15.83~\text{Hz}$.

\item Mechanical decoherence rate due to photon scattering from the nanosphere: $\Gamma_m^s/2 \pi=76.5 \times 10^{-3}~\text{Hz}$.

{\item Mechanical decoherence rate due to photon scattering from the black body radiation: $\Gamma_{m}^{BB}/2 \pi=5.8 \times 10^{-7}~\text{Hz}$.}

\end{itemize}

\end{document}